\documentclass[review]{elsarticle}

\usepackage{amsmath} 
\usepackage{lineno,hyperref}
\modulolinenumbers[5]

\usepackage{tikz}
\usepackage{pgfplots}
\usepackage{subcaption} 

\journal{Journal of \LaTeX\ Templates}

\usepackage{numcompress}

\begin{document}
\begin{frontmatter}

\title{The Tersoff potential for extreme environment }


\author{Youhwan Jo}
\author{Taeyeon Kim}
\author{Byeongchan Lee\corref{mycorrespondingauthor}}

\address{Kyung Hee University}
\address{1732 Deogyeong-daero, Yongin, Gyeonggi 17104, Republic of Korea}
\cortext[mycorrespondingauthor]{Corresponding author}
\ead{airbc@khu.ac.kr}




\begin{abstract}
Not finished yet.
\end{abstract}

\begin{keyword}
\texttt{Molecular dynamics, Tersoff Potential, Silicon}
\end{keyword}

\end{frontmatter}


\section{Introduction}

Ever since the late 20th century, molecular dynamics (MD) has been a useful technique for developing a new material or understanding a phenomenon in the microscopical level. Typically, quantum molecular dynamics or ab-initio methods where interactions between electrons are directly concerned has demonstrated it's reliability through many research results with the accurate reproductions of the qualitative nature of materials.   

Modern engineering technologies occasionally demand a thorough study on an extraordinary topic, as an example, nuclear fusion technologies requires an exceptional material, which can withstand severe neutron flux at the same time is thermally and mechanically stable. Because it is very difficult to arrange an experiment for such topic, proving or examining a possible candidate material by an experiment is very limited. Molecular dynamics is a perfect complement for that situation. Actually, many researchers have used molecular dynamics to see the effect of irradiations from such as, a high energy ion, in the way of primary knock-on atom (PKA) simulations. Although the available order of time or spatial dimension makes ab-initio methods impractical for such simulations, classical molecular dynamics itself is still quite attractive to see directly the evolution of cascades and the thermal response of materials in the atomistic level.

The major problem of molecular dynamics (from now on, molecular dynamics typically stands for the classical molecular dynamics.) is a reliability of an interatomic potential. Especially, achieving a good transferability is very important because the transferability is a direct measure for the reliability of simulation results related to a topic which is difficult to check by experiments, in other words, the transferability represents the validity of molecular dynamics upon where it is really meaningful. Unfortunately, it is very difficult to achieve a good transferability. 

In this paper, a novel modification of the Tersoff potential~\cite{tersoff1988new} for Si is presented. The modification improves the transferability of the Tersoff potential for liquid states without the change of original parameters and with no alteration of bulk properties. Also, the modification introduces a correction term for high-pressure states. Si, especially SiC composites are a prospect candidates for a structural compartment for nuclear fusion reactors, for example, flow channel inserts (FCI) in dual coolant lead lithium (DCLL) blankets. The modification is meaningful considering that by high energy irradiations local liquid structures and unstable high-pressure manifolds may occur, therefore an interatomic potential must have an acceptable reliability on high thermal/pressure situations to simulate such phenomenon.  Particularly, in the modification, a novel screening function replaces a radial cutoff function and a bond order function is slightly changed. Also, a repulsive energy function is replaced by a correction function within a specific pair distance. All features of the modification are thoroughly explained in the section 3.  
 
\section{The Tersoff Potential}

The Tersoff potential is named after J. Tersoff who originally introduced the potential at 1988 \cite{tersoff1988new}. Many variations of the Tersoff potential have been offered since the introduction, nonetheless, the core of the potential is still intact. The mathematical expression of the original format is like below:   

\begin{equation}
\begin{aligned}
\label{eq:1}
&V_{ij} = f_{c}(r_{ij})[f_{r}(r_{ij}) + b_{ij}f_{a}(r_{ij})],\\
\end{aligned}
\end{equation}

\begin{equation}
\begin{aligned}
\label{eq:morse}
&f_{r}(r_{ij}) = A\exp{(-\lambda_{1}r_{ij})},\\
&f_{a}(r_{ij}) = -B\exp{(-\lambda_{2}r_{ij})},
\end{aligned}
\end{equation}\

\noindent where $f_{r}$ and $f_{a}$ are a repulsive and an attractive energy term respectively, specifically $f_{a}$ is multiplied by $b_{ij}$, a bond order function, whose value depends on the atomic environment of a pair as the below manner.

\begin{equation}
\begin{aligned}
\label{eq:bij}
&b_{ij} = (1+\beta^n\zeta_{ij}^n)^{-1/2n},
\end{aligned}
\end{equation}\

\noindent with

\begin{equation}
\begin{aligned}
\label{eq:zetaij}
&\zeta_{ij} =  \sum_{k{\neq}i,j}\zeta_{ijk},\\
&\zeta_{ijk} = f_{c}(r_{ik})g(\theta_{ijk})k(r_{ij},r_{ik}),\\
&g(\theta_{ijk})=1+(c/d)^2 - c^2/[d^2+(h-\cos{\theta_{ijk}})],\\
&k(r_{ij},r_{ik})=\exp{[\lambda_{3}^\kappa(r_{ij}-r_{ik})^\kappa]}.
\end{aligned}
\end{equation}\

\noindent The inverse of the bond order function is effectively equivalent to a coordination number \cite{tersoff1988new}. Because the cohesive energy of the Tersoff potential changes with respect to an atomic environment, typically the Tersoff potential is classified as environment dependent interatomic potentials (EDIP) or by accentuating the bond order function, bond order potentials (BOP). 

\begin{equation}
\begin{aligned}
\label{eq:fc}
&{f_{c}(r) =}\begin{cases}
	1, & r < R-D, \\
	\frac{1}{2}-\frac{1}{2}\sin{[\frac{\pi}{2}(r-R)/D]}, & R-D < r < R+D, \\
	0, & r > R+D,
	\end{cases}
\end{aligned}
\end{equation}\

$f_{c}(r)$ is a radial cutoff function and has a transition in the range of 2D around R. The maximum cutoff $R+D$ is typically a distance between 1NN and 2NN of interesting structures in favor of the bond order function, which estimates the coordination number that represents the number of 1NN. As an example, for diamond cubic (DC) Si, J. Tersoff offered $R+D = 3.0$ \AA{} \cite{tersoff1988revised}. 

Many studies with the Tersoff potential have presented promising results regardless its relatively simple form. The potential predicts relatively well properties related to bulk states such as elastic constants, phonon frequencies \cite{balamane1992comparative}. For simulating bulk phase transformations, the Tersoff potential is also competitive among other potentials \cite{balamane1992comparative}.  Also, it is noteworthy that the Tersoff potential predicts well defect formation energies, although related to surfaces the potential may give a wrong result except (100) surface for Si \cite{balamane1992comparative}. Threshold displacement energies, which are important quantities considering irradiation simulations like the primary knock-on atoms (PKA) method, the Tersoff potential reproduces an anisotropic nature nonetheless absolute values are underestimated \cite{holmstrom2008threshold, miller1994displacement}.

Probably one of the most notable weaknesses of the Tersoff potential is simulating non-crystalline states, especially liquid states.  
It is well known that the Tersoff potential fairly overestimates the melting temperature. The amount of the overestimation is slightly different in various reports, nonetheless it is more than about 700K~\cite{tersoff1988revised, cook1993comparison, ishimaru1996tm, yoo2004tm}. Moreover, the pair correlation of liquid states is inaccurate that is characterized by a notable dropout after the first peak~\cite{tersoff1988revised, cook1993comparison}. Such disparities are the sign of misleading results when simulating liquid states. 


If considering nuclear fusion environments, another problem emerges. A neutron which comes out as a result of fusion reactions has 14.1 MeV energy and possibly transfers some part of the energy to an atom by a collision. The amount of the transferred energy ranges over few to several hundreds keV depending on various factors. Obviously it exceeds the conventional frame of molecular dynamics considering that an interatomic potential is typically made for simulating thermally equilibrium states (moreover, the majority of training sets also physical properties related to those states.). Simply, it cannot be guaranteed that an energetic response to an exceptionally high energy input would be realistic. This problem is actually universal for almost every interatomic potentials, consequently many researchers have tried to find an answer from other solutions. One example is the two temperature model (TTM), which implements an electronic temperature in addition to the atomic temperature. The TTM simulates ion-electron interactions by employing a Langevin type thermostat by which imaginary electrons thermally interact with an atom \cite{duffy2006TTM, rutherford2007effect}. If the kinetic energy of an atom is greater than a specified level, a friction term is changed to include an electronic stopping power therefore it is possible to embody an energy loss by inelastic collisions with electrons reasonably \cite{duffy2006TTM, rutherford2007effect}.     

The TTM is fundamentally adding a new feature of physics into a simulation method, the reliability of a simulation is still doubtable because basically atoms follow trajectories formed by an interatomic potential. Generally, the short-range interaction of most interatomic potentials is considered unrealistic, many researchers used the Ziegler-Biersack-Littmark potential (ZBL) which represents the Coulomb repulsion \cite{ziegler1985stopping} as an alternative by joining or coupling the ZBL potential with an interatomic potential. Specifically, for the Tersoff potential, \Citeauthor{devanathan1998-tZBL} combined the ZBL potential (parameter values are tweaked to match a density functional theory calculation.) with the Tersoff potential using the Fermi function~\cite{devanathan1998-tZBL} and \Citeauthor{belko2003-tZBL} replaced the two-body part of the Tersoff potential with the ZBL potential~\cite{belko2003-tZBL}.

\section{In Need of a Modification}

\begin{figure}\centering
	\begin{subfigure}[b]{0.5\textwidth}
		\centering
		\resizebox{\linewidth}{!}{
		 \includegraphics[scale=1]{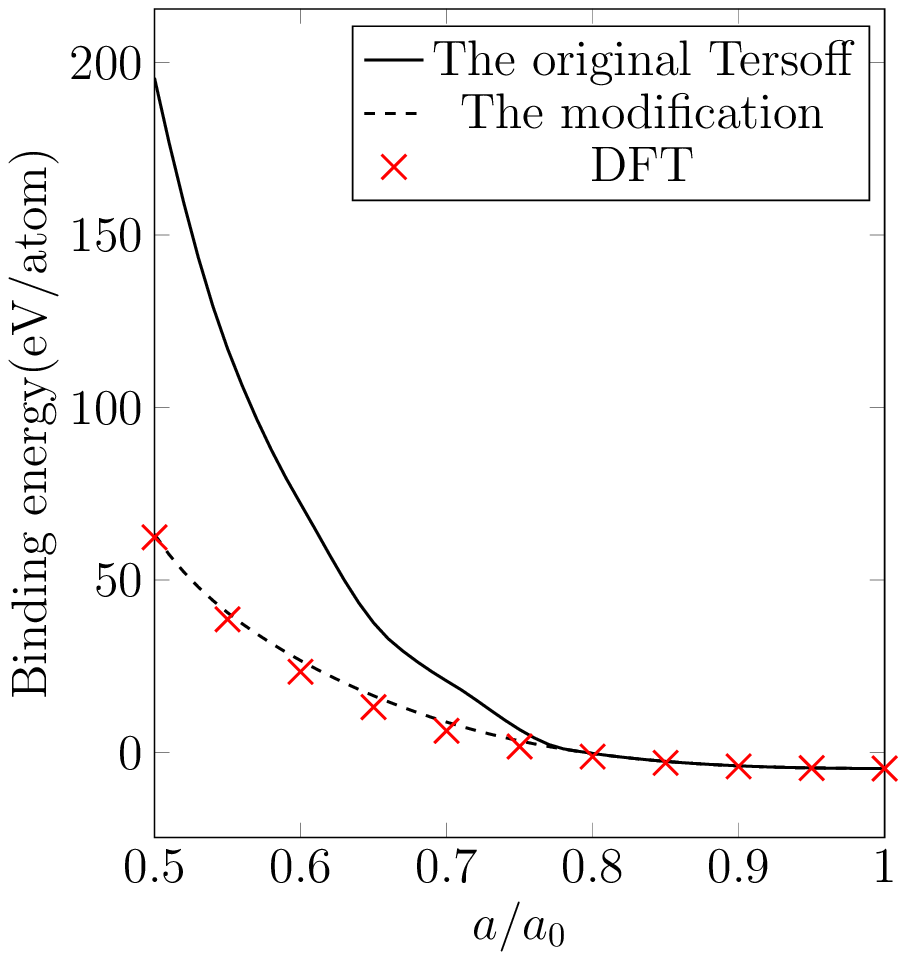}
		 }
                 \caption{Binding energy profile}
                 \label{fig:fittingResult-a}
	\end{subfigure}%
	\begin{subfigure}[b]{0.5\textwidth}
		\centering
		\resizebox{\linewidth}{!}{
			\includegraphics[scale=1]{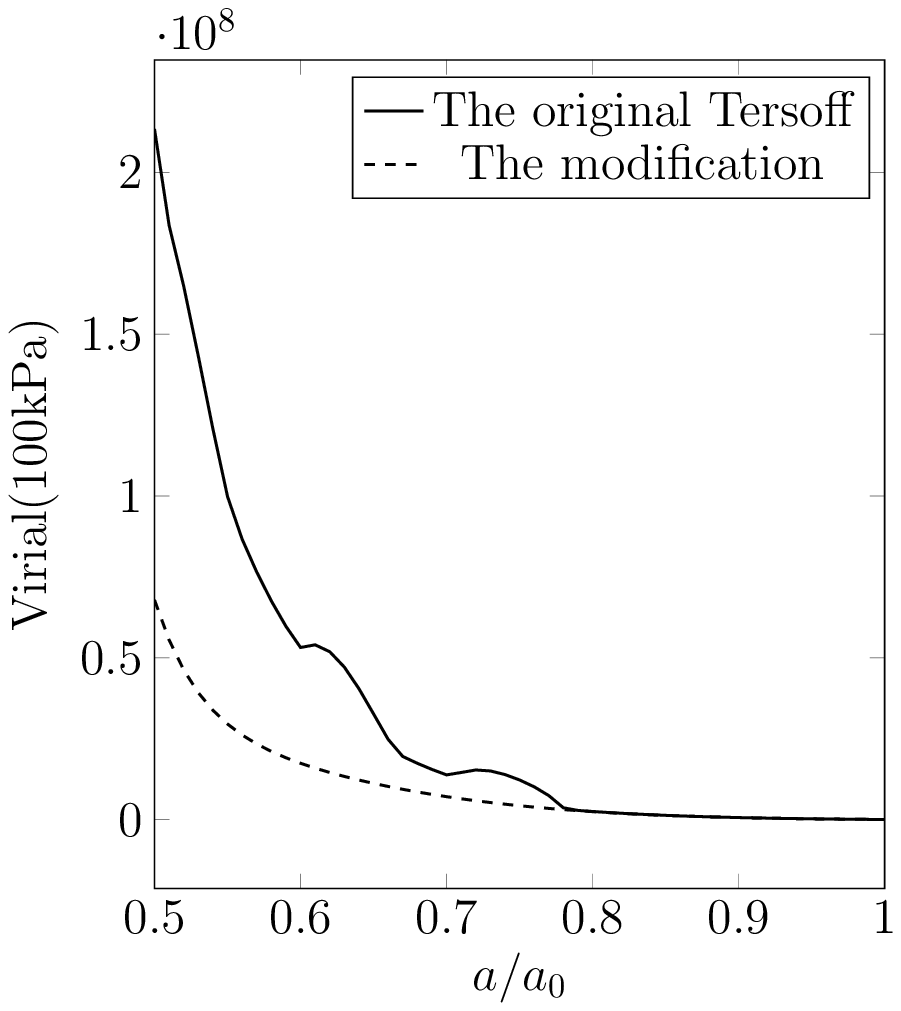}
			}
                 \caption{Virial pressure profile}
                 \label{fig:fittingResult-b}
	\end{subfigure}

\caption{Profiles for the compressed DC-Si. The original Tersoff stands for the original function form of the potential with parameter values in Table.~\ref{table:parameterValues}. $a_{0}$ is a lattice constant for the ground state.}
\label{fig:fittingResult}
\end{figure}

The purpose of this research is improving the reliability of the Tersoff potential in the context of high thermal/pressure situations. The best way to achieve such goal is finding a new training way which presents a wonderful new parameter set by which all problems are eliminated. However, through numerous failures, we concluded that it requires any form of a functional modification to improve the Tersoff potential. In this section, reasons for such conclusion will be explained.

\begin{figure}\centering
\includegraphics[scale=1]{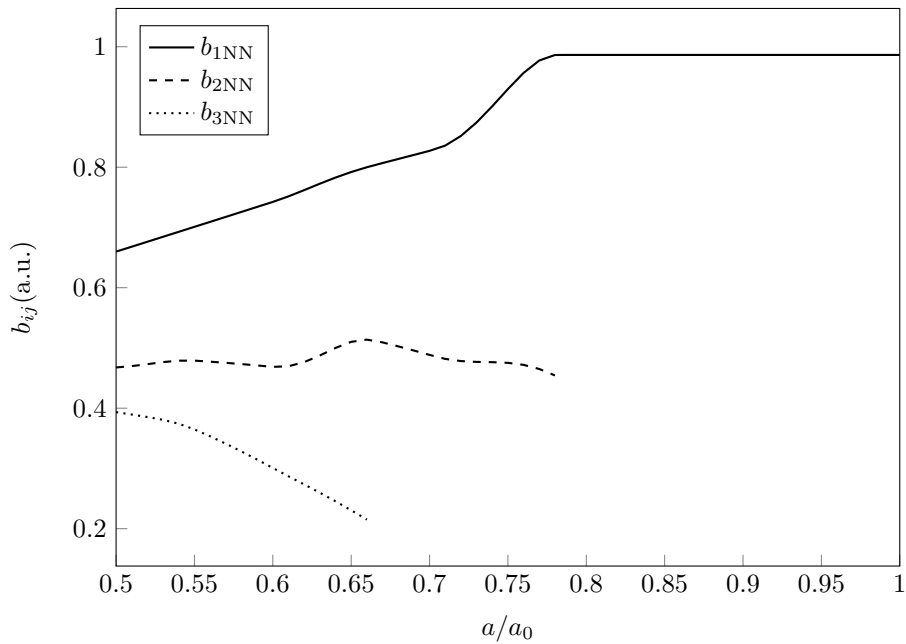}
\caption{$b_{ij}$ for each NN during the compression of DC-Si.}\centering
\label{fig:bij_original}
\end{figure}

\subsection{The Repulsive Energy Function, $f_{r}(r)$}

The short-range interactions of the Tersoff potential may become realistic by adding a high-pressure state data into a training set although it probably leads the unavoidable sacrifice of a reproducibility for other entries. An alternative function which replaces the original function when a pair distance is short is a reasonable solution to earn a reliability with no sacrifice on the reproducibility. Another justification for an alternative function comes from the Morse-style energy function of the Tersoff potential. As described in Eq.~\ref{eq:morse}, the cohesive energy will be determined as the combination of two exponential functions and this particular form converges to a constant value as a pair distance becomes zero. It is obviously not physically plausible and even though a new parameter set successfully achieves both an accuracy to the high-pressure state data and an acceptable degradation of the reproducibility, in the extrapolation region where the data cannot cover, a desirable dependency on a pair distance may not be acquired due to the inherent disability of the function.

Like what we mentioned in the section 2, the ZBL potential is a possible candidate for the alternative function. Although we hesitated simply to use the ZBL potential. Although it is hard to specify which physics will be dominant on the short-range interactions and more fundamentally it may be inappropriate to express that physics with an interatomic potential, it seems that the restriction of the $1/r$ character of the Coulomb interaction is inappropriate to capture the nature of short-range interactions. A polynomial whose degree is not specified probably is the best option in that sense. In the section 4, a correction function, which is basically a polynomial function will be introduced as the alternative function.

Then, what reference would be legitimate as representing a high-pressure state? We decided to use a cohesive energy profile of compressed diamond cubic Si by a DFT calculation. Although the energy coverage of the profile is less than about 60 eV, as it will be shown in the section 4, the extrapolated curve of a polynomial correction function rapidly surges. 

\subsection{The Radial Cutoff Function, $f_{c}(r)$}

After a few tryouts, we realized that fitting the Tersoff potential to the cohesive energy profile from the DFT calculation is almost impossible even with an additional correction function. In Fig. \ref{fig:fittingResult}, the profile of the Tersoff potential deviates from the DFT result and moreover has fluctuating biases which do not exist in the DFT result. The main reason for those disparities is the cutoff function of the potential. As the structure is compressed, at some point, a pair distance for any further neighbor group (2NN, 3NN and so on.) becomes shorter than the cutoff radius. Although conceptually the Tersoff potential counts interactions between 1NN, a simple radial cutoff function like $f_{c}$ (Eq.\ref{eq:fc}) counts all atomic pairs as valid interactions whose distances are shorter than a cutoff radius. Additional atoms which come within a radial cutoff suddenly change a bond order value for indigenous pairs (an example can be seen in Fig.\ref{fig:bij_original}.) and apparently the change of the bond order value has transitions (fluctuations) due to the cutoff function inside $b_{ij}$.  As a result, the attractive energy is substantially decreasing also fluctuating by newly introduced pairs, on the other hand, the repulsive energy is still intact because it only depends on a pair distance. Consequently, the energy profile deviates as the repulsive energy will be dominant more as the structure is compressed more.

Moreover, the cutoff function is a major cause for the incompetent transferability to l-Si. More precisely, a small value for a cutoff transition $D$ is problematic (typically, $D$ is 0.2 \cite{tersoff1988new} or  0.15 \cite{tersoff1988revised} \AA.) because any atomic interaction in between ${R-D}$ and ${R+D}$ fades with a slope that is proportional to $1/2D$. Any atomic pair whose pair length belongs to such cutoff transition range is energetically unstable, in other words, the cutoff function acts like an artificial energy barrier. Such effect can be directly seen in the pair correlation of l-Si with the Tersoff potential in multiple references \cite{tersoff1988revised, cook1993comparison}. Pair correlations suddenly drop nearby the cutoff transition range because a pair within that range is unstable and statistically unfavorable. 

Actually, it is not a rare case that a cutoff function (specifically, a 2-body radial cutoff function) invokes an undesirable result. \Citeauthor{pastewka2008_1} reported that cutoff functions are responsible to the should-be-brittle-but-ductile nature of bond order potentials (BOP) \cite{pastewka2008_1, pastewka2013_2}.  A similar phenomenon also happened with the modified embedded atom potential (MEAM) \cite{ko2014origin}. It is worth to notice that such fallacy is not merely the problem of a short cutoff. \Citeauthor{aghajamali2018unphysical} reported that the Tersoff potential with an extended cutoff shows an unphysical nucleation of carbon \cite{aghajamali2018unphysical}. 

Evidently a better cutoff mechanism is necessary. A new mechanism must be possible to discriminate 1NN from others and also a radial cutoff for the new mechanism must be sufficiently long therefore any artificiality during atomic transition events would be as small as possible. An obvious conclusion is replacing the cutoff function with a screening function. The benefit of a screening function over a cutoff function is already presented by other groups \cite{pastewka2008_1, pastewka2013_2, kumagai2009screening}.

\begin{figure}\centering
\includegraphics[scale=1]{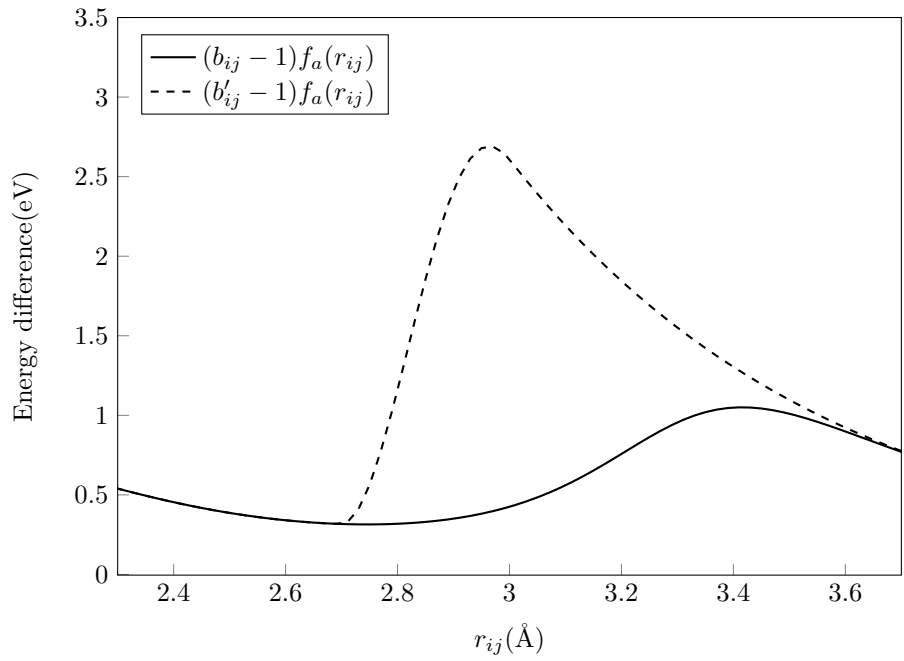}
\caption{The effect of the bond order function and the cutoff function as changing $r_{ij}$ when $r_{ik}=2.35\textrm{\AA}$, $\theta_{ijk}=109.5^{\circ}$ are fixed. $b'_{ij} = f_{c}(r_{ij})b_{ij}$ and the parameter set Si(C) from Ref.~\citenum{tersoff1988revised} is used.}\centering
\label{fig:bij_as_barrier}
\end{figure}

\subsection{The Bond Order Function, $b_{ij}$}

For a bond order potential like the Tersoff potential, one thing must be considered before replacing a cutoff function. Actually, a cutoff function in BOP not only trims an atomic interaction out but also defines the scope of the bond order \cite{aghajamali2018unphysical}. As mentioned before, by replacing the cutoff function we also want to increase a cutoff radius, then the scope of the bond order is quantitatively and qualitatively changed. 

A bond order contribution by a neighboring atom is defined as $\zeta_{ijk}$ in {Eq.~\ref{eq:zetaij}} and any cutoff or screening function can be used for disregarding the contribution of an atom. Within $\zeta_{ijk}$, $k(r_{ij},r_{ik})$ is a quite interesting feature. \Citeauthor{tersoff1988new} mentioned the meaning of $k(r_{ij},r_{ik})$ by setting $\kappa$ as 3: the effect of atoms to each other is different with respect to a relative pair length \cite{tersoff1988new}. It is quite reasonable that a shorter pair is less affected by a longer pair, nonetheless, the penalty on a longer pair by a shorter pair in the form of the bond order function $b_{ij}$ is actually tremendous. Although in the Tersoff potential that penalty is hard to catch because a short radial cutoff trims a pair interaction before such penalty prevails. {Fig.~\ref{fig:bij_as_barrier}} shows the dominance of $f_{c}(r_{ij})$ over $b_{ij}$ when the radial cutoff is sufficiently short. As seen in {Fig.~\ref{fig:bij_as_barrier}} if the cutoff function is totally removed, the bond order function $b_{ij}$ will behave like another artificial energy barrier at some point. 

It is inappropriate to judge that Tersoff's idea about $k(r_{ij},r_{ik})$ is right or wrong because $k(r_{ij},r_{ik})$ is empirically decided rather having a physical basis. Nonetheless, it is possible to say that whether $k(r_{ij},r_{ik})$ is good or not for a specific purpose. The research of \citet{kumagai2007development} is illuminating for this matter. \Citeauthor{kumagai2007development} offered the modified version of the Tersoff potential (Tersoff-MOD) and successfully produces a melting temperature that is almost equal to the experimental result with this potential \cite{kumagai2007development}. Atypically, the Tersoff-MOD has a longer cutoff than the original and sets $\kappa$ as 1 unlike 3 of the original.  Apparently the energy barrier from the cutoff function is reduced and at the same time the penalty from the bond order function is weakened. We guess that a strong correlation exists between the accurate melting temperature and such changes of parameter values (although there are other contributions because the form of $g(\theta_{ijk})$ and $f_{c}(r_{ij})$ are also changed.). With the Tersoff-MOD, the sudden dropout of the pair correlation of l-Si also become weaker than the original Tersoff although it is still notable \cite{pun2017optimized}. 

We urge that the penalty from $k(r_{ij},r_{ik})$ should be totally removed to improve the transferability to l-Si at mostly. Changing the parameters $\lambda_{3}$ and $\kappa$ in $k(r_{ij},r_{ik})$ simply decides how strongly and equivalently the penalty from $k(r_{ij},r_{ik})$ is. By setting $\lambda_{3} = 0$, $k(r_{ij},r_{ik})$ can be inactive nonetheless it seems not a good solution considering another degradation may happen (The l-Si density does not match with the experiment value when $\lambda_{3} = 0$.) \cite{cook1993comparison}. Therefore it is necessary to change the form of $k(r_{ij},r_{ik})$.









\section{The Modification for the Tersoff Potential}

We offer a novel modification for the Tersoff potential by which aforementioned problems of the Tersoff potential are eliminated or at least mitigated. The modification has three main parts: the screening function, the correction function, and the refashioned zeta.  

\subsection{The Screening Function}

\begin{equation}
\begin{aligned}
&{I_{ij} =}\begin{cases}
	I_{1}, 	&	 {P_{ij} \geq1,} \\
	I_{1}[1-(1-P_{ij})^{I_{2}}]^{I_{3}}, 	&	 {1 > P_{ij} > 0,} \\
	0, 	&	 {P_{ij} \leq 0.}
	\end{cases}
\end{aligned}
\end{equation}\

The screening function ($I_{ij}$) is simply a smooth function from the MEAM potential of \citet{baskes1997smooth}. The function replaces the original cutoff function, therefore it is multiplied to pair energy functions and also be a part of $\zeta_{ijk}$. The form of the potential after the replacement is like below:

\begin{equation}
\begin{aligned}
&V_{ij} = I_{ij}[f_{r}(r_{ij}) + b_{ij}f_{a}(r_{ij})],\\
&b_{ij} = (1+\beta^n\zeta_{ij}^n)^{-1/2n},\\
&\zeta_{ij} = \sum_{k{\neq}i,j} \zeta_{ijk},\\
&\zeta_{ijk} = I_{ik}g(\theta_{ijk})k(r_{ij},r_{ik}),\\
&k(r_{ij},r_{ik})=\exp{[-\lambda_{3}^2(r_{ij}-r_{ik})^2]}.
\label{eq:themodification}
\end{aligned}
\end{equation}\

\noindent An input for the screening function, $P_{ij}$ (the screening factor) represents the mechanism of screening. Although, \Citeauthor{baskes1997smooth} takes the ratio between axes of an ellipse made of three atoms as a screening factor \cite{baskes1997smooth}, in this modification, we offer another way to define a screening factor. Our screening factor has the following form:       

\begin{equation}
\begin{aligned}
&{P_{ij} =}\begin{cases}
	\dfrac{S_{ij}^2}{S_{max}^2}, & S_{ij} > 0, \\
	0, & S_{ij} \leq 0, \\
	\end{cases}\\
&S_{ij} = S_{0} + \sum_{k{\neq}i,j} S_{ijk}, \\
&S_{0} = -S_{1}[r_{ij}/(R+D)]^2 + S_{2},\\
&S'_{ijk} = (r_{ik}-r_{ij})[r_{ik}-(R+D)]^4,\\
&S_{ijk} = S'_{ijk}/{(r_{ik}^2r_{jk}^3)}.
\label{eq:screeningFactor}
\end{aligned}
\end{equation}\

Our whole idea about the new screening factor is starting from a simple fact: the easiest way to identify each NN is comparing pair distances. The function $S'_{ijk}$ is the epitome of such thought. $S'_{ijk}$ is a polynomial function whose value is zero when $r_{ik}$ equals to $r_{ij}$ or $(R+D)$. It bisects any atomic configuration into two regions on the behalf of a specific pair $r_{ij}$. The boundary of two regions is a surface where $r=r_{ij}$, consequently the two regions are an inner sphere $(r < r_{ij})$ with a negative $S'_{ijk}$ and an outer sphere $(r_{ij} < r < R+D)$ with a positive $S'_{ijk}$. Also, the shape of $S'_{ijk}$ depends on $r_{ij}$, therefore the effect of $S'_{ijk}$ is different for each NN. The function $S_{ijk}$ is simply the non-dimensionalized version of $S'_{ijk}$ nonetheless the shape of a boundary is deformed by a dividing factor (in this case, $r_{ik}^2r_{jk}^3$). The summation of $S_{ijk}$ gives a distinct value for each NN ($S_{\textrm{1NN}}$, $S_{\textrm{2NN}}$ and so on.). In other words, it is possible to identify each NN with this function. In our screening mechanism, an atomic pair competes in a pair length with others within a cutoff and for each competition, the pair will take a positive or negative value ($S_{ijk}$). By analogy to this nature, we named the function $S_{ijk}$ as the three-body score function.

\begin{figure}\centering
\includegraphics[scale=1]{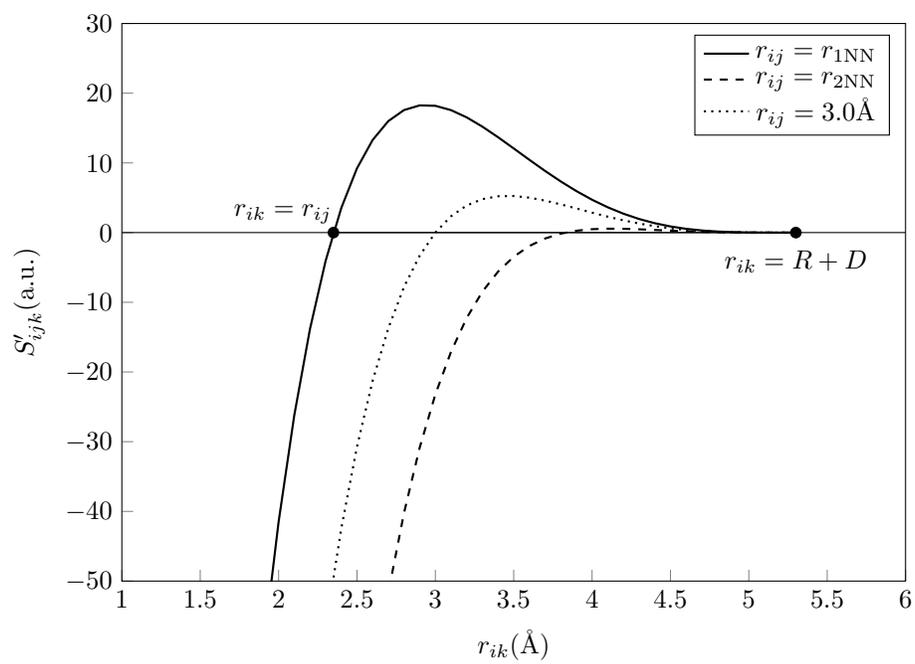}
\caption{The dependency of $S'_{ijk}$ to $r_{ij}$. When $R+D=5.3$ {\AA}.}\centering
\label{fig:the_sijk_primitive}
\end{figure}

It is necessary to decide what value of the score would be a fully screened point (where $I_{ij}$ becomes 0.). Any value between $S_{\textrm{1NN}}$ and $S_{\textrm{2NN}}$ would be sufficient to screen 2NN out. Nevertheless we set $S_{ij} = 0$ as the fully screened point considering that in any crystalline structure with any length of a cutoff, 1NN would have a score value which equals or larger than 0. Then the screening factor $P_{ij}$ is simply a squared score value normalized by a saturation factor $S_{max}$. Also the squared form of $S_{ij}$ makes sure that the derivatives of $P_{ij}$ would be zero when a pair is fully screened, therefore the screening mechanism will not violate the conservation of energy. For a specific atom ${i}$, the shape of a surface where $S_{ij} = 0$ is dependent on an atomic configuration near the atom $i$. That surface encloses the atom ${i}$ like a candy wrapper and an interaction with an atom outside this surface will be fully screened (It is conceptually an extension of the boundary made by $S'_{ijk}$.). The radial cutoff ($R+D$) only decides a possible neighboring atom ${j}$ which might interact with the atom ${i}$, whether a pair is screened (trimmed out) or not is now dependent on the relative position of an atom ${j}$ and the atomic environment of the atom ${i}$.

\begin{figure}\centering
\includegraphics[scale=1]{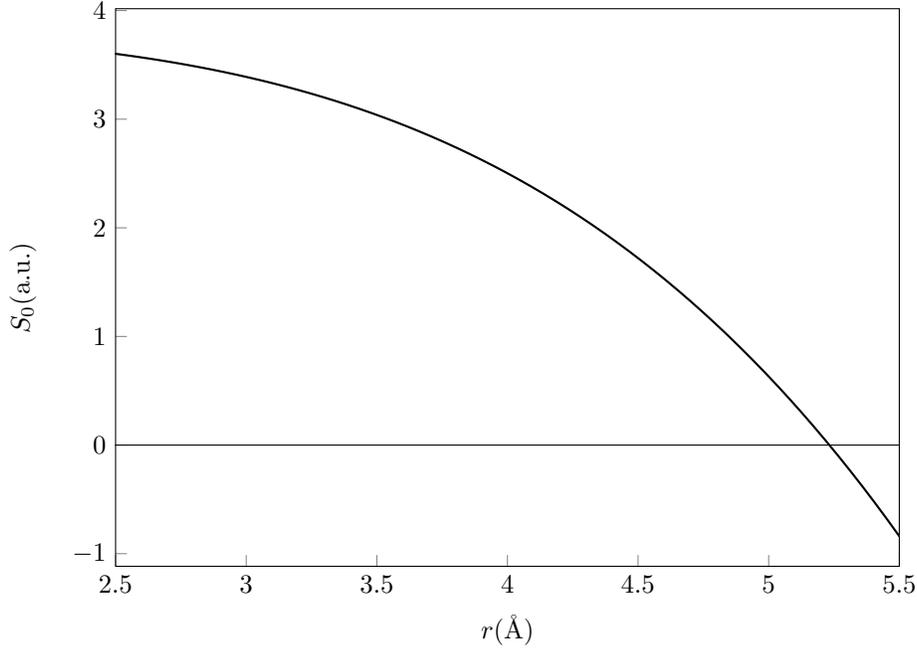}
\caption{The single pair score $S_{0}$.}\centering
\label{fig:the_s0}
\end{figure}

However, only with the three-body score function, the screening mechanism is incomplete because the screening factor cannot be defined unless at least three atoms exist. Therefore it is necessary to use an additional two-body cutoff function for two-body situations. Instead, we introduced the pair score function $S_{0}$. The pair score function give an additional score to an atomic pair with respect to their pair length. If such additional score is a negative value nearby a radial cutoff, any interaction with an atom near the cutoff will be screened out. For an atomic dimer, the pair score typically do what the cutoff function do in the original Tersoff potential. In that case, adjusting parameters $S_{1}$ and $S_{2}$ is equivalent to change the cutoff parameters $R$ and $D$. For general cases, the pair score function changes the shape of a non-screened region from the summation of $S_{ijk}$. 

Values for the additional parameters of the screening function ($I_{1}$, $I_{2}$ and others) are handpicked and the normalization factor in $S_{ijk}$, polynomial orders for $S'_{ijk}$ and $S_{0}$ are empirically decided. Parameter values are given in Table.~\ref{table:parameterValues}. Conditions that we used to decide additional parameter values are (a) except 1NN, all pair interactions are fully screened until $\epsilon = 0.5$, (b) in DC, an effective distance of screening is longer than $(r_{\textrm{1NN}}+r_{\textrm{2NN}})/2$. About the condition (b), it is simply checked where an atom is screened or not by moving it away from an original tetrahedral point along the direction of the pair.

\begin{table}[]
\centering
\resizebox{\linewidth}{!}{
    \begin{tabular}{llllll}
    \hline\hline
    \multicolumn{2}{c}{Screening} &  \multicolumn{2}{c}{Correction} &  \multicolumn{2}{c}{Original}\\
    \hline
    $I_{1}$	&	1.0	&	$A_{0}~(\textrm{eV})$	&	-140579.439268257	&	$A~(\textrm{eV})$	&	1832.050054\\ 
    $I_{2}$	&	3.0	&	$A_{1}~(\textrm{eV})$	&	680.033744052508	&	$B~(\textrm{eV})$	&	471.1103049\\
    $I_{3}$	&	4.0	& 	$A_{2}~(\textrm{eV})$ 	&	-0.757496473019405	& 	$\lambda_{1}~({\textrm{\AA}}^{-1})$ &	2.4712291\\
    $S_{1}$	&	4.0	& 	$\mu~({\textrm{\AA}}^{-1})$	&	5.622176179440607	&	$\lambda_{2}~({\textrm{\AA}}^{-1})$  &	1.738128648\\
    $S_{2}$	&	3.8	& 	$\omega$ &	5.43945312500000	&	$\beta$  & 2.0299E-06\\
    $S_{max}$	&	2.53	& $r_{\textrm{replace}}~({\textrm{\AA}})$	&	2.139601	&	$n$	&	0.799302308\\
    $r_{cut}~({\textrm{\AA}})$	&	5.3	&	{}	&		&	$c$	&	179340.388293655\\
    {}	&	{}	&	{}	&		&	$d$	&	24.42304604 \\
    {}	&	{}	&	{}	&		&	$h$	&	-0.457616628330656 \\
    {}	&	{}	&	{}	&		&	$\lambda_{3}~(\textrm{\AA}^{-1})$	&	1.1\\
    \hline
    \end{tabular}
}
\caption{Parameter values for the modification. $r_{cut} = R+D$.}\centering
\label{table:parameterValues}
\end{table}

\subsection{The Refashioned zeta}

The refashioned zeta is already presented in Eq.~\ref{eq:themodification}. We changed $k(r_{ij},r_{ik})$ from $\exp{[\lambda_{3}^3(r_{ij}-r_{ik})^3]}$, which \Citeauthor{tersoff1988revised} suggested \cite{tersoff1988revised} to $\exp{[-\lambda_{3}^2(r_{ij}-r_{ik})^2]}$. Setting $\kappa = 2$ is tried before in our previous work \cite{kim2013science}, but a negative sign is a new feature of this work. By $\kappa = 2$, the penalty from $k(r_{ij},r_{ik})$ becomes equivalent and additionally the negative sign make the penalty totally disappear.

\subsection{The correction function}

\begin{equation}
\begin{aligned}
&f_{\textrm{correction}}(r_{ij}) = A_{0}\exp({-{\mu}r_{ij}}) + \dfrac{A_{1}}{r_{ij}^{\omega}} + A_{2},
\end{aligned}
\label{eq:thecorrectionfunction}
\end{equation}

\begin{equation}
\begin{aligned}
&{f_{r} =}\begin{cases}
	A\exp({-\lambda_{1}r_{ij}}), 	&	 {r_{ij} \geq r_{\textrm{replace}},} \\
	f_{correction}, 	&	 {r_{ij} < r_{\textrm{replace}}.}
	\end{cases}
\end{aligned}
\label{eq:thecorrection_condition}
\end{equation}\	

\noindent As depicted in Eq.~\ref{eq:thecorrection_condition}, the correction function replaces the repulsive term $f_{r}(r_{ij})$ of the original Tersoff potential within a specific pair distance, $r_{\textrm{replace}}$. The exponential part in the correction function typically minimizes a discrepancy between the DFT data and the modification and the rest part of the correction, a polynomial function makes sure that the pair energy diverges to the positive infinite as a pair length become shorter.

As presented in Eq.~\ref{eq:thecorrectionfunction}, the correction function has the five parameters but four of them are exclusively decided to acquire a C3 continuity with the original repulsive term at $r_{\textrm{replace}}$ (functions which depend $A$, $\lambda_{1}$, $r_{\textrm{replace}}$, and the left parameter decide the value of the four parameters.). Therefore, only one additional parameter is used for fitting the DFT data. Parameter values are given in Table.~\ref{table:parameterValues}.

\section{The Results}

\begin{figure}\centering
\includegraphics[scale=1]{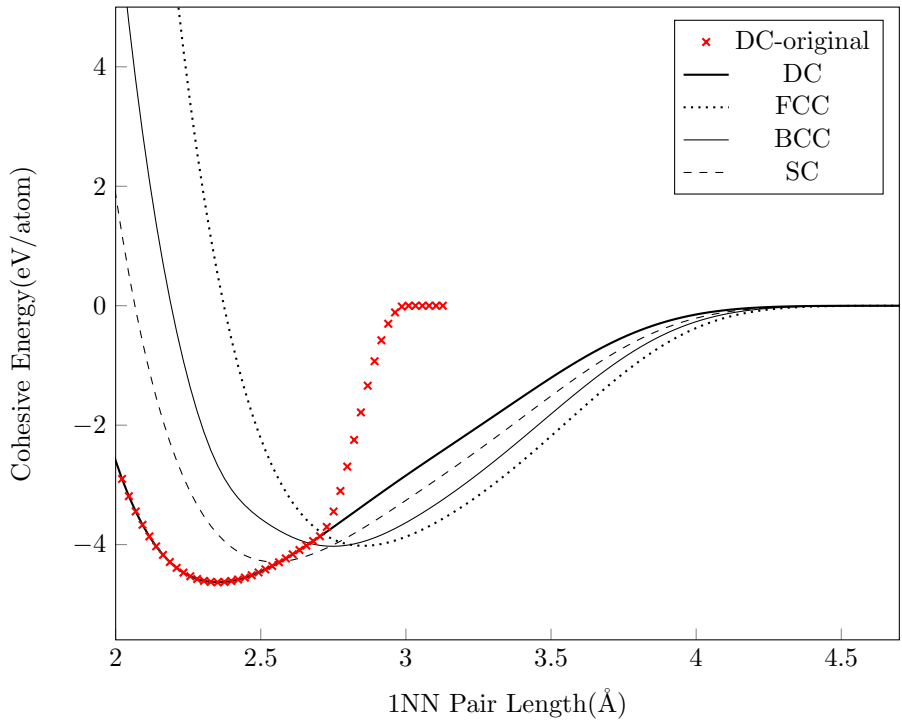}
\caption{The phase stability with the modification. The DC-original stands for the original form of the Tersoff potential with $R=2.85$ {\AA} and $D=0.15$ {\AA}.}\centering
\label{fig:phaseStability}
\end{figure}

By replacing the cutoff function with the screening function, the modification removes an artificiality from the cutoff function in the tensile domain. In this work, $r_{\textrm{cut}} = 5.3$~\AA, this value is long enough to pass an inflection point of energy when parameter values in Table.~\ref{table:parameterValues} are used. Actually, it is fairly long that a binding energy around 5.3~$\textrm{\AA}$ is almost negligible as you may see in Fig.~\ref{fig:phaseStability}. In the tensile domain, the screening factor mostly depends on $S_{0}$ therefore 1NN maintain interactions until $S_{0}$ returns a negative value.

\begin{figure}\centering
\includegraphics[scale=1]{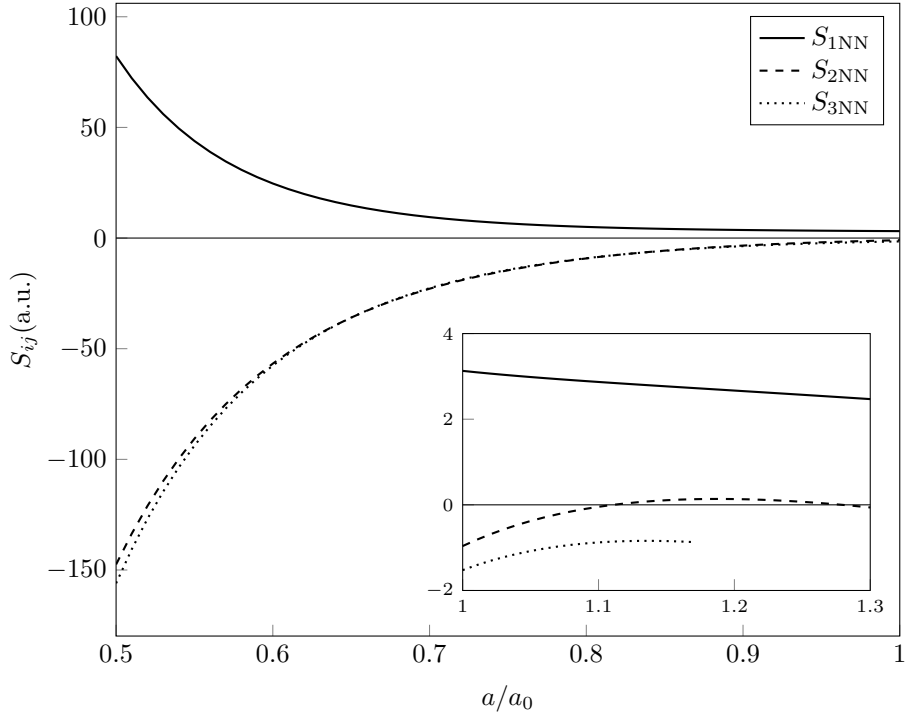}
\caption{$S_{ij}$ values for each NN during the compression of DC-Si.}\centering
\label{fig:score_compressive}
\end{figure}

Fig.~\ref{fig:score_compressive} shows that score values are negative except $S_{\textrm{1NN}}$ in the compressive domain, consequently as seen in Fig.~\ref{fig:fittingResult}, the energy deviation and fluctuations due to 2NN and 3NN are totally removed. Although the change of score values is monotonic in the range of our consideration, it might be reversed like $S_{\textrm{2NN}}$ in the inset figure of Fig.~\ref{fig:score_compressive}. Because $S_{ij}$ is transformed to $P_{ij}$ like as in Eq.~\ref{eq:screeningFactor}, the positive values of $S_{\textrm{2NN}}$ do not have a noticeable effect in the tensile domain.

\begin{figure}\centering
\includegraphics[scale=1]{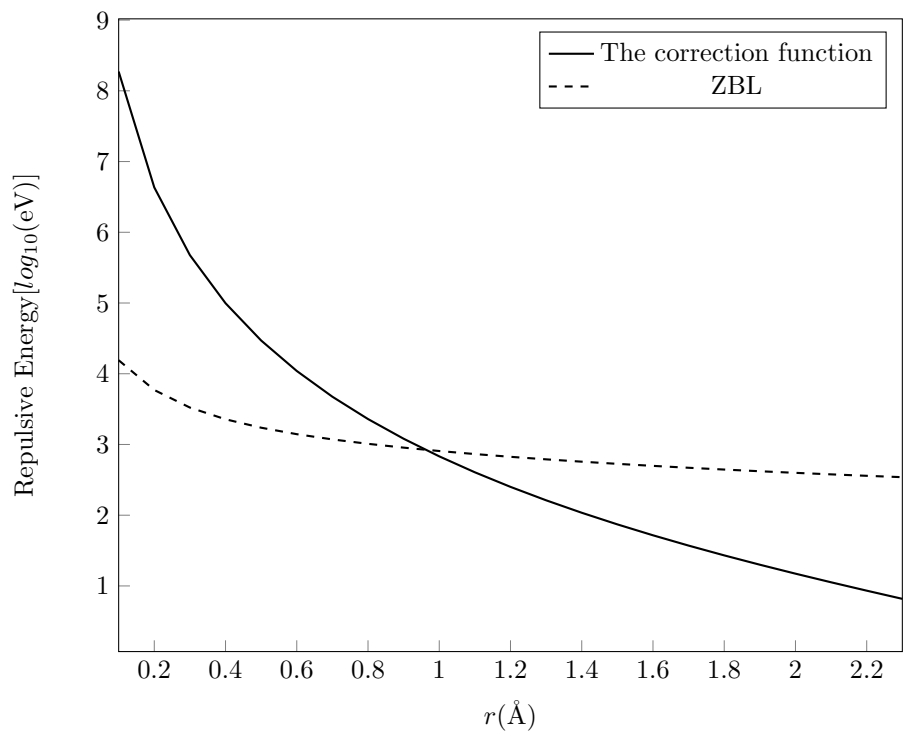}
\caption{The difference between the correction function and the ZBL potential.}\centering
\label{fig:zbl&thecorrection}
\end{figure}

Even if the screening function excludes atomic interactions with 2NN and further atoms, a mismatch happens with the original Tersoff potential. By the correction function in Eq.~\ref{eq:thecorrectionfunction} that mismatch is reduced. Also, the polynomial part of the correction function guarantees that energy diverges to positive infinite unlike the exponential energy term of the original Tersoff. It is also worth to check a difference between the correction function and the ZBL potential. The polynomial degree of the correction function is a parameter that can be determined and the resultant degree, $\omega$ is about 5.43 (Table.~\ref{table:parameterValues}) in this work. As seen in Fig.~\ref{fig:zbl&thecorrection} the difference is huge, the correction function covers few hundreds keV order of energy in a reasonable pair length unlike the $1/r$ nature limits the energetic coverage of the ZBL potential.  

\begin{figure}\centering
\includegraphics[scale=1]{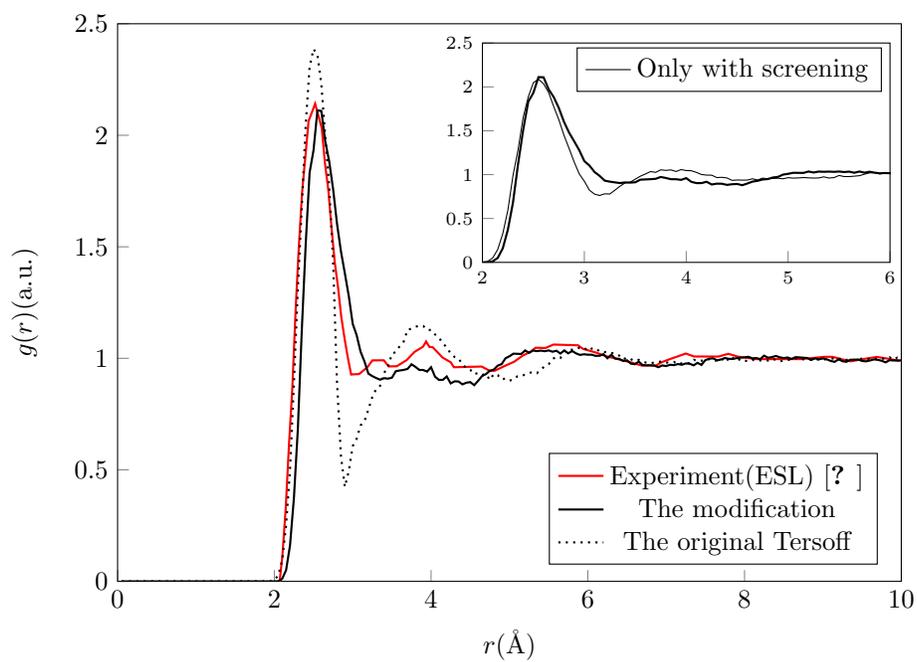}
\caption{Pair correlations for l-Si. Temperatures for each entry are different.}\centering
\label{fig:grChange}
\end{figure}

\begin{figure}\centering
\includegraphics[scale=1]{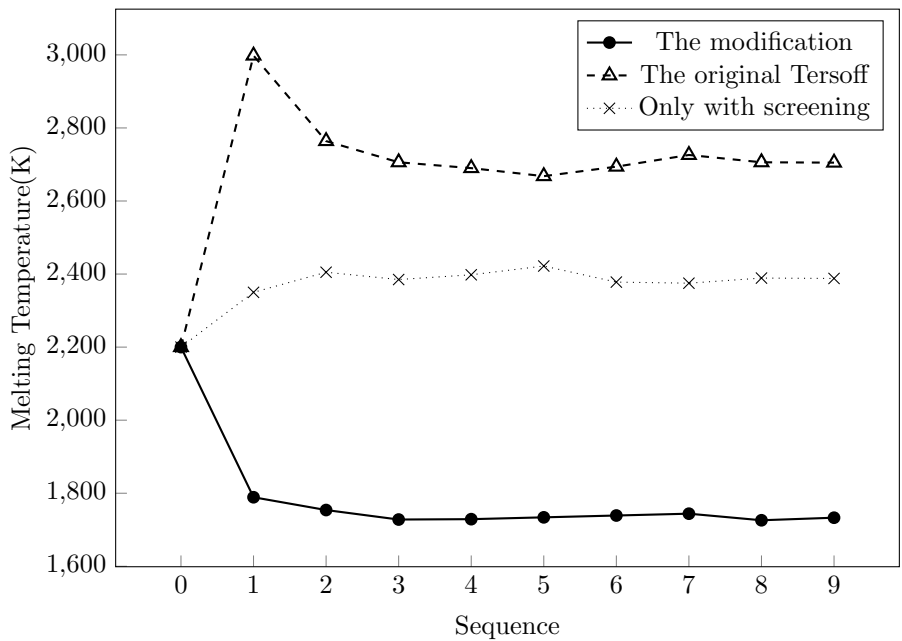}
\caption{Coexistence simulation sequences for each function forms. All parameters are same except $(R+D)$ for every cases.}\centering
\label{fig:meltingTemperature}
\end{figure}

An improved transferability to liquid states with the modification is quite impressive. The signs of the incompetent transferability of the original Tersoff potential, the overestimated melting temperature and the sudden dropout in the pair correlation are considerably corrected. Fig.~\ref{fig:grChange} shows the change of the pair correlation. The liquid structure of the experiment result is earned by an electrostatic levitation technique (ESL) and further information can be seen in Ref.~\cite{kim2005situ}. The improved transferability is the result of the entire modification. As you can see in Fig.~\ref{fig:grChange}, the dropout is still noticeable if only the screening function is applied. We estimated a melting temperature by coexistence methods, with the modification, the average value of estimated melting temperatures is about 1741 K. Like the case of pair correlations, with only the screening function, the improvement becomes less significant (Fig.~\ref{fig:meltingTemperature}).

Some properties of liquid states also changed by the modification. We found that an l-Si density become more accurate (the modification: 2.578, the original: 2.52).

\begin{figure}\centering
\includegraphics[scale=1]{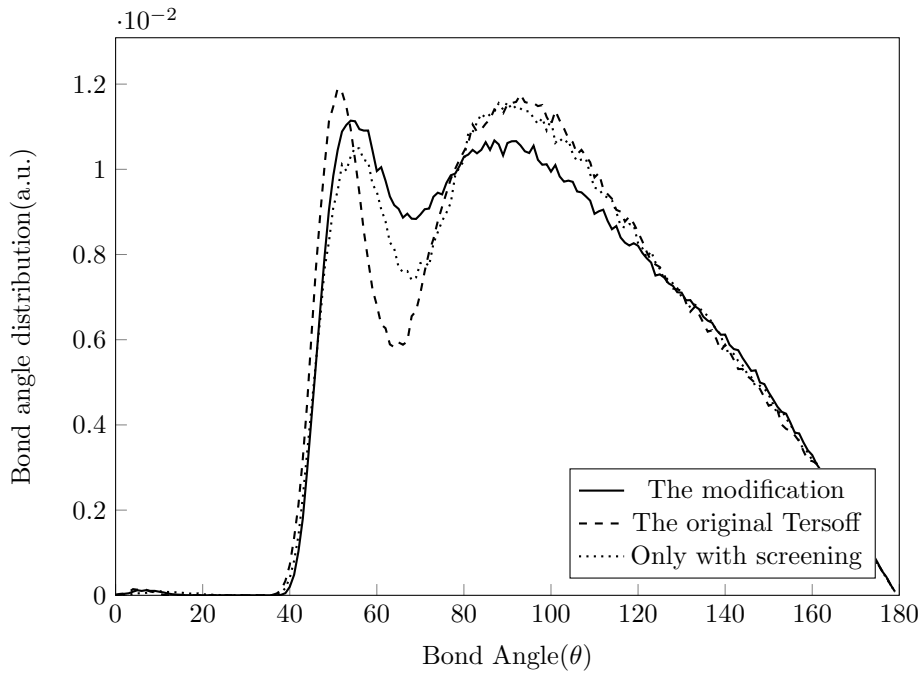}
\caption{Bond angle distributions for l-Si. Temperatures for each entry are different. 3.3~\AA is the maximum range of calculations.}\centering
\label{fig:baChange}
\end{figure}

\section{The Discussion}

Our screening mechanism based on the score of an atom has unique characters. First, the screening mechanism does not need a radial cutoff function that is similar to the work of \citet{pastewka2013_2}. Although we did not test our modification, this character is a great merit for fracture and sputtering simulations because an artificial stress increasing due to a radial cutoff function is removed. Second, the screening mechanism depends on an atomic scale. As mentioned in the section 4, the three-body score function $S_{ijk}$ changes with respect to $r_{ij}$. Consequently whether an atom is screened or not would be different even with a same structure if a scale is different like 2NN in Fig.~\ref{fig:score_compressive}. Third, an additive nature. A score value $S_{ij}$ is the result of summations through many functions in the manner of Eq.~\ref{eq:screeningFactor}. The single presence of an atom is not decisive like the MEAM screening function where an atom which satisfies $C_{ij} < C_{\textrm{min}}$ precludes the screening by a multiplication \cite{baskes1997smooth}. It needs much research to clarify a difference between the additive approach and the productive way, nonetheless a hint could be found in the report of \Citet{ryu2008comparison}. They mentioned that by reducing $C_{min}$ some thermal properties is improved with examples in their own work \cite{ryu2008comparison} and the second nearest neighbor MEAM (2NN-MEAM) \cite{lee2000second}. A possible deduction is that a geometrical interpretation of screening (like an ellipse.) might be a constraint for a certain angular structure. 

The improvement on the pair correlation is encouraging, still a noteworthy difference exists between the result with the modification and the experiment. It is a subtle peak around 3.3 {\AA} of the ESL curve in Fig.~{\ref{fig:grChange}}. As far as we know, such subtle small peak is only catchable by the experiment and the Stillinger-Weber potential (SW) \cite{cook1993comparison, stillinger1986chemical}. The comparison between the Tersoff and the SW potential might reveal the secret of the subtle peak.  

\begin{figure}\centering
\includegraphics[scale=1]{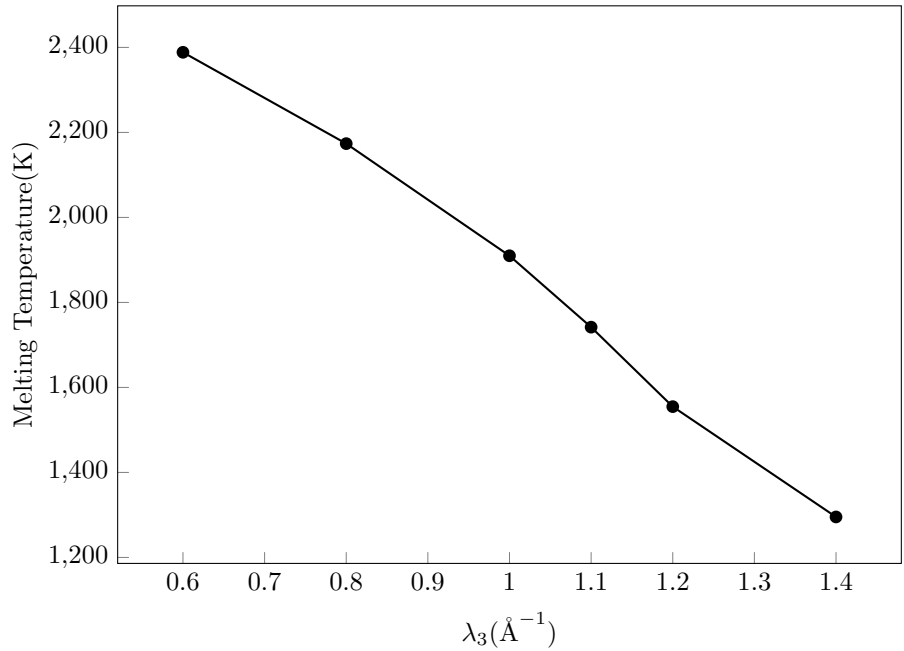}
\caption{The relation between melting temperatures and values of ${\lambda_{3}}$.}\centering
\label{fig:l3dependency}
\end{figure}

With the refashioned zeta, we found a strong correlation between a melting temperature and $\lambda_{3}$ as you can see in Fig.~\ref{fig:l3dependency}. The correlation is actually obvious considering that the energy penalty by the form of $b_{ij}$ is decreased as increasing the value of $\lambda_{3}$. Functionally, the new form of $k(r_{ij},r_{ik})$ in Eq.~\ref{eq:themodification} defines an effective range of ${b_{ij}}$. Therefore in some situation, the bond order can be separated with the screening function; although a neighboring kth atom is not fully screened, the effect of the ik-pair on ${b_{ij}}$ can be totally disregarded by $k(r_{ij},r_{ik})$.

Although

\section{The Conclusion}

\section*{References}

\bibliography{mybibfile}

\end{document}